\documentclass[aps,pra,reprint,floatfix,superscriptaddress]{revtex4-2}
\usepackage{amsmath, amssymb}
\usepackage{graphicx} 
\usepackage{physics}
\usepackage{MnSymbol}
\usepackage{color}
\usepackage{xcolor}
\usepackage[shortlabels]{enumitem}

\usepackage{dcolumn}
\usepackage{multirow}
\usepackage{hyperref}

\newcommand{\nmel}{E}

\begin{document}

\title{Enhanced Compressive Threshold Quantum State Tomography for Qudit Systems}

\author{Giovanni Garberoglio}
\author{Maurizio Dapor}
\affiliation{European Centre for Theoretical Studies in Nuclear Physics and Related Areas (FBK-ECT*), Strada delle Tabarelle 286, Trento (Italy).}

\author{Diego Maragnano}
\author{Marco Liscidini}
\affiliation{Department of Physics, University of Pavia, Pavia (Italy).}

\author{Daniele Binosi}
\affiliation{European Centre for Theoretical Studies in Nuclear Physics and Related Areas (FBK-ECT*), Strada delle Tabarelle 286, Trento (Italy).}

\date{\today}

\begin{abstract}
We propose an efficient quantum state tomography method inspired by compressed sensing and threshold quantum state tomography that can drastically reduce the number of measurement settings to reconstruct the density matrix of an $N$-qudit system. We validate our algorithm with simulations on IBMQ and demonstrate the efficient and accurate reconstruction of $N\leq7$ qubit systems, reproducing GHZ, $W$, and random states with ${\cal O}(1)$, ${\cal O}(N^2)$, and ${\cal O}(N)$ settings.
\end{abstract}

\maketitle

\section{Introduction}
Quantum state tomography (QST) is a fundamental technique for reconstructing the density matrix $\rho$ of a quantum system by measuring a sufficiently large number of observables. In the standard approach, known as full QST (fQST), one must measure the expectation value of at least $d^{2N}$ observables to reconstruct the density matrix of a system composed of \textit{N} qu\textit{d}its, \textit{i.e.}, \textit{N} \textit{d}-level carriers of information.  
The number of measurements to perform fQST scales polynomially with the dimension of the Hilbert space, which itself scales exponentially with the number of qudits. These elements make fQST challenging even for relatively small size systems. 
Strategies implementing fQST select a sufficiently large set of measurements to gather all the information needed to reconstruct the density matrix.

Mutually unbiased bases (MUBs) represent a common approach to fQST \cite{klimov2008optimal, dambrosio2013test, giovannini2013characterization, klimov2013optimal, rohling2013tomography, yuan2016quantum, yuan2022qutrit, wang2024classical}. 
Measurements using MUBs are optimal for QST as they provide a minimal, maximally independent, and informative set of measurements \cite{ivonovic1981geometrical, wootters1989optimal, lawrence2002mutually, adamson2010improving}. However, constructing MUBs explicitly for large Hilbert spaces becomes increasingly difficult, and their existence is not guaranteed if the dimension is not a power of a prime.
(Over)complete sets of projective measurements are widely used to perform fQST too, especially on photonic systems that exploit the degrees of freedom of polarization and frequency \cite{james_2001, renes2004symmetric, zhu2014quantum, bent2015experimental, lu2022bayesian}. 
Finally, one can carry out fQST by measuring the expectation value of measurement settings, \textit{i.e.}, products of the generators of SU($d$). The number of required settings is in this case $d^{2N}$, with an equivalent number of projective measurements equal to $d^{3N}$ \cite{thew2002qudit, haffner2005scalable, pont2024high}.

Several methods have been developed to optimize fQST. Among these, compressed sensing QST can reconstruct the density matrix with fewer measurement settings than fQST. However, the reduction in the number of settings is significant only for pure states, and the settings are chosen randomly, without optimization \cite{gross_2010, gross2011recovering}. There have been theoretical studies on the complexity of compressed sensing QST, adaptive variations, and experimental implementations \cite{flammia2012quantum, liu2012experimental, tonolini2014reconstructing, steffens2017experimentally, ahn2019adaptive, ahn2019adaptive2, badveli2020compressed}. 

Recently, threshold Quantum State Tomography (tQST) was introduced for qubit systems ($d=2$)~\cite{Maragnano23,tQST}. This method aims to reduce the number of required projective measurements, without any assumption about the state being reconstructed. 
The algorithm leverages on the inequality $\lvert \rho_{ij} \rvert \leq \mathfrak{r}_{ij} = \sqrt{\rho_{ii} \rho_{jj}} $ to introduce a threshold \( t \) and focus on the matrix elements \( (i,j) \) for which \( \mathfrak{r}_{ij} \geq t \). 
In the limit \( t \to 0 \), the tQST procedure requires \( 2^{2N} \) projective measurements, and is equivalent to fQST. However, for states characterized by a sparse density matrix, tQST enables to reconstruct the density matrix with fewer projective measurements. 

In this work, we introduce a method applicable to any dimension $d$ by combining tQST and measurement settings made out of SU(\textit{d}) generators. We leverage the core idea of tQST and compressed sensing to drastically reduce the number of measurement settings without compromising the quality of the reconstructed state. For these reasons, we call this method Enhanced Compressive Threshold Quantum State Tomography (ECT-QST). This approach can be applied to quantum systems of any dimension without any assumption on the state, giving the possibility to reconstruct states that are out of reach for present methods. The work is organized as follows. First, we detail the steps of the ECT-QST algorithm. We then demonstrate its efficiency for systems of up to 7 qubits, first in the noiseless case and then on the IBMQ supercondcting platform. Finally, we discuss further research directions and draw our conclusions.

\section{The ECT-QST algorithm }
This section outlines the ECT-QST algorithm, focusing on $N$-qudit systems unless otherwise specified.

\subsection{\label{sec:1-qudit}1-qudit observables}

Let us assume the $d^2 - 1$ operators $\sigma^{(k)}$, corresponding to the generators of SU($d$), can be measured on each qudit. Of these, $d-1$ operators span the Maximal Abelian Subgroup (MAS); they all have the same eigenvectors which define the computational basis. Thus, we only need to retain a single element from this set, or, without loss of generality, replace them with the $d$-dimensional identity, $\mathbb{I}_d$. In either cases, we will denote the MAS representative as $\sigma^{(0)}$.

The remaining $d(d-1)$ operators are represented by Hermitian matrices with only two non-zero elements located in off-diagonal positions. Half of these matrices are real and the other half are purely imaginary. They can be indexed by the position of their non-zero upper-triangular element, starting with the real-valued matrices followed by the imaginary ones. For example, for $d=2$ there are three qubit observables, which correspond to the usual SU(2) Pauli matrices: $\sigma^{(0)} = \sigma_z$ (the MAS generator), $\sigma^{(1)} = \sigma_x$ and $\sigma^{(2)} = \sigma_y$. 
For $d=3$ the 1-qutrit operators are provided instead by the 8 Gell-Mann matrices, which are the SU(3) analogous of the Pauli-matrices above. The real operators read
\begin{align}
	\sigma^{(1)} &= 
\begin{pmatrix}
0 & 1 & 0 \\
1 & 0 & 0 \\
0 & 0 & 0
\end{pmatrix};&
	\sigma^{(2)} &= 
\begin{pmatrix}
0 & 0 & 1 \\
0 & 0 & 0 \\
1 & 0 & 0
\end{pmatrix};&
	\sigma^{(3)} &= 
\begin{pmatrix}
0 & 0 & 0 \\
0 & 0 & 1 \\
0 & 1 & 0
\end{pmatrix},
\end{align}
whereas the corresponding imaginary operators are given by
\begin{align}
	\sigma^{(4)} &= 
\begin{pmatrix}
0 & i & 0 \\
-i & 0 & 0 \\
0 & 0 & 0
\end{pmatrix};&
	\sigma^{(5)} &= 
\begin{pmatrix}
0 & 0 & i \\
0 & 0 & 0 \\
-i & 0 & 0
\end{pmatrix};&
	\sigma^{(6)} &= 
\begin{pmatrix}
0 & 0 & 0 \\
0 & 0 & i \\
0 & -i & 0
\end{pmatrix}.
\end{align}
Finally, the SU(3) MAS comprises two generators:
\begin{align}
\sigma^{(7)} &= 
\begin{pmatrix}
1 & 0 & 0 \\
0 & -1 & 0 \\
0 & 0 & 0
\end{pmatrix};&
	\sigma^{(8)} = &= \frac{1}{\sqrt{3}}
\begin{pmatrix}
1 & 0 & 0 \\
0 & 1 & 0 \\
0 & 0 & -2
\end{pmatrix}.
\end{align}

\subsection{\label{sec:choice_of_settings}Choice of settings}

As in tQST~~\cite{Maragnano23,tQST}, the algorithm  starts with measuring the $d^N$ diagonal elements $\{ \rho_{ii} \}$ of the density matrix. This involves measuring all $d^N$ states formed as products of the eigenvectors of $\sigma^{(0)}$ on each qudit, and corresponds to the setting $s^{(00\cdots0)}=\otimes_{r=1}^N \sigma^{(0)}$. 
After this first measurement, one uses a suitably chosen threshold $t$ to select the matrix elements $\rho_{ij}$ for which $\mathfrak{r}_{ij} \ge t$.

The choice of threshold is a crucial step in this procedure, as it determines the amount of resources required for tomography. In general, the optimal value of $t$  depends on the noise level of the system under consideration. Ideally, one would like to choose a threshold that is small enough to include the most relevant elements of the density matrix, yet large enough to exclude those elements that are most likely affected by noise and do not contribute meaningful information about the quantum state being measured. In Appendix~\ref{app:threshold-IBMQ} we describe the approach that we have used to fix $t$ in the experiments on the IBMQ platform. 
In the absence of information guiding the choice of $t$, considerations based on measures of sparsity, like the Gini index~\cite{Gini1914} discussed in Appendix~\ref{app:threshold-Gini}, can also provide useful guidance.

Next, we seek for a set $S_t$ of measurement settings of the form $s^{(K)}=s^{(k_1k_2\cdots k_N)} = \otimes_{r=1}^N \sigma^{(k_r)}$ that provide optimal information on the selected matrix elements. Here, $K = \left( k_1, k_2, \dots , k_N \right)$ is a multi-index; each index $k_r$ ranges from 0 to $d(d-1)$ and describes which observable is measured on the $r$-th qudit.  As the expectation value of $s^{(K)}$ on $\rho$ is given by
\begin{align}
 \langle s^{(K)} \rangle 
 &= 2 \sum_{i \leq j} \left[ 
 \mathrm{Re}\ s^{(K)}_{ij}\ \mathrm{Re}\ \rho_{ij}+ \mathrm{Im}\ s^{(K)}_{ij}\ \mathrm{Im}\ \rho_{ij}
\right],
\label{eq:info}   
\end{align}
it follows that a sufficient condition for $s^{(K)}$ to provide information on the real (imaginary) part of $\rho_{ij}$ is that the real (imaginary) part of $s^{(K)}_{ij}$ is non-zero. From the definition of $s^{(K)}$ one has
\begin{align}
s^{(K)}_{ij} &= s^{(K)}_{i^d_1 i^d_2 \cdots i^d_N, j^d_1 j^d_2 \cdots j^d_N} =  \sigma^{(k_1)}_{i^d_1, j^d_1}
\sigma^{(k_2)}_{i^d_2, j^d_2}
\cdots\sigma^{(k_N)}_{i^d_N, j^d_N}, 
\label{eq:S_base_d}
\end{align}
with $i^d_r$ the $r$-th digit of the base-$d$ representation of $i$ (and similarly for $j^d_r$).
Equation~(\ref{eq:S_base_d}) shows that for $s^{(K)}_{ij}$ to be non-zero, \textit{all} the matrix elements of the single-qudit operators on the right-hand side of Eq.~(\ref{eq:S_base_d}) must be non-zero. 

To find a setting providing information about the real part of $\rho_{ij}$, we digit-wise compare the base-$d$ representations of $i$ and $j$ and choose:
$\sigma^{(0)}$ if $i^d_r = j^d_r$; and
$\sigma^{(k_\ell)}$, corresponding to the unique real generator with a non-zero element at $(i^d_r,j^d_r)$, if $i^d_r \neq j^d_r$.
Correspondingly, to gain information about the imaginary part of the same element, it is sufficient to change an odd number of single-qudit operators where $i^d_r \neq j^d_r$ to the corresponding imaginary ones. 
We choose to change only one single-qudit operator, the one for which $i_r \neq j_r$ for the smallest value of $r$. Thus, to gain information on
	the imaginary part of $\rho_{ij}$, we replace the first real single-qudit operator $\sigma^{(k_r)}$ different from $\sigma^{(0)}$ with the corresponding $\sigma^{(k_r+d(d-1)/2)}$. 
For example, in the case of $N=4$ qubits ($d=2$) the setting associated to the real part of the matrix element $i=4=0100^2$ and $j=13 = 1101^2$, is given by $s^{(1001)}=\sigma_{x}\otimes\sigma_{z}\otimes\sigma_{z}\otimes\sigma_{x}$; the corresponding imaginary part is measured instead by the setting $s^{(2001)}=\sigma_{y}\otimes\sigma_{z}\otimes\sigma_{z}\otimes\sigma_{x}$. For $N=3$ qutrits ($d=3$) the setting measuring the real part of the matrix element $i=12=110^3$ and $j=23=212^3$ is $s^{(302)}=\sigma^{(3)}\otimes\sigma^{(0)}\otimes\sigma^{(2)}$ with the corresponding imaginary part measured by  $s^{(602)}=\sigma^{(6)}\otimes\sigma^{(0)}\otimes\sigma^{(2)}$.

In the $t \to 0$ limit, one measures all the elements of $\rho$, and the algorithm would require measuring at most $2 \times [d(d-1)/2+1]^N$ settings, resulting in ${\cal O}(d^{3N}/2^N)$ projective measurements.
For sparse density matrices with $\nmel$ 
matrix elements to be determined, this procedure identifies ${\cal O}(\nmel)$ settings, corresponding to ${\cal O}(\nmel d^N)$ projective measurements. Consequently, it is highly likely that some settings provide information on multiple matrix elements, making them redundant.

\subsection{\label{sec:pruning}Pruning}

In order to find the minimum number of settings that still enable reconstructing the density matrix under consideration, it is useful to introduce operators that provide information on the real or imaginary part of a given density matrix element. Indicating with $m$ a multi index in the form $(i,j,{\mathcal R})$ for real elements or $(i,j,{\mathcal I})$ for imaginary ones, the operators $O_m$ fulfilling
\begin{equation}
    \mathrm{tr}(O_m \rho) = 
    \left\{
    \begin{array}{ll}
       \mathrm{Re}(\rho_{ij})  & \mathrm{if\ } m=(i,j,{\mathcal R})  \\
       \mathrm{Im}(\rho_{ij})  & \mathrm{if\ } m=(i,j,\, {\mathcal I}) 
    \end{array}
    \right.,
    \label{eq:Om}
\end{equation}
are such that the only non-zero elements of $O_m$ are $(i,j)$ and $(j,i)$, with their values being both $1/2$ if $m=(i,j,{\mathcal R})$ or $i/2$ and $-i/2$ if $m=(i,j,{\mathcal I})$. Since a setting $s=s^{(K)}$ prescribes a sequence of $N$ observables $\sigma^{(k_r)}\in$ SU$(d)$, it corresponds to a separable measurement basis $|\phi^{(s)}_n\rangle = \otimes_{r=1}^N | \varphi^{(k_r)}_{n^d_r} \rangle$, with $| \varphi^{(k_r)}_{n^d_r} \rangle$ eigenstates of $\sigma^{(k_r)}$.
Defining $P^{(s)}_{n} = | \phi^{(s)}_n\rangle\langle \phi^{(s)}_n|$, we next construct the rank-3 tensor
\begin{align}
	A^{(s)}_{mn} &= \langle \phi^{(s)}_n | O_m | \phi^{(s)}_n \rangle
	= \mathrm{tr}\left(O_m P^{(s)}_{n}\right),
 \label{eq:A}
\end{align}
which indicates how well the $n$-th element of the measurement basis corresponding to setting $s$ overlaps with the matrix element $m$. If $A^{(s)}_{mn} = 0$, measuring the expectation value of the projector $P^{(s)}_{n}$ will not provide any information on the matrix element $m$.
Correspondingly, the matrix 
\begin{align}
	C_{sm} &= \sum_n \left|A^{(s)}_{mn}\right|^2\ \forall\ s\in S_t,
\end{align}
quantifies how well measurements in the basis of the setting $s$ provide information on the matrix element $m$. We will call $C_{sm}$ the overlap of setting $s$ with the matrix element $m$.
Denoting by $|X|$ the cardinality of the set $X$, $C_{sm}$ is a $|S_t| \times (2 \nmel)$ matrix, with $\lvert S_t \rvert \le 2E$.

Consider then the maximum overlap on the matrix element $m$ among the settings $s \in S_t$, that is
\begin{align}
	\beta_m = \mathrm{max}_{s} ~ C_{sm}, 
\end{align}
which surely exists but ought not to correspond to a unique $s$. In general, $\sum_s C_{sm} \gg \beta_m$, indicating that many projectors $P^{(s)}_{n}$ in different settings $s\in S_t$ provide information on the same matrix element $m$; we are therefore looking for a subset $S_t' \subset S_t$ so that $\sum_{s \in S_t'} C_{s m} \approx \beta_m$, so that we aim at reducing the number of settings while keeping the total overlap with the matrix element $m$ as large as the maximum overlap among the settings identified in Sect.~II B above.

The simplest way to find such a subset is through a {\it greedy} algorithm, which attempts to find a globally optimal solution by making locally optimal choices. Start by selecting the setting $s_1$ such that $C_{s_1 m}$ has the fewest elements equal to zero. This ensures that $s_1$ provides information on the maximum number of matrix elements of interest.
Add $s_1$ to $S_t'$ and remove it from $S_t$. 
Find the setting $s_2 \in S_t$ with the next fewest zeros in $C_{s_2 m}$, excluding the columns $m'$ for which the target value has already been reached, that is the columns $m'$ such that $\sum_{s' \in S_t'} C_{s' m'} \geq \beta_{m'}$. Continue the process until the condition $\sum_{s' \in S'_t} C_{s'm} \geq \beta_m$ is satisfied for all $m$.

By the end of this procedure, we have a small{\it er} (close to, if not, the small{\it est}) set of settings, $S'_t$, whose elements have an overlap with the matrix element $m$ (that is, $\sum_{s' \in S_t'} C_{s' m}$) as large as the maximum overlap among the original settings (that is, $\beta_m$).

\subsection{\label{sec:sorting}Sorting}

The  settings in \( S'_t \) found after pruning are not equally important for the reconstruction of \( \rho \). To address this, we assign a weight \( w_s = \sum_{m} C_{sm} \mathfrak{r}_m \) to each setting, which reflects the total overlap of $s$ on the matrix elements that are to be determined. We then sort the settings in descending order based on their weight and measure them progressively, prioritizing the most significant ones.  

An implementation of ECT-QST using the Python language is available online.~\cite{github}

\begin{figure*}[!t]
	\centering
	\includegraphics[scale=0.265]{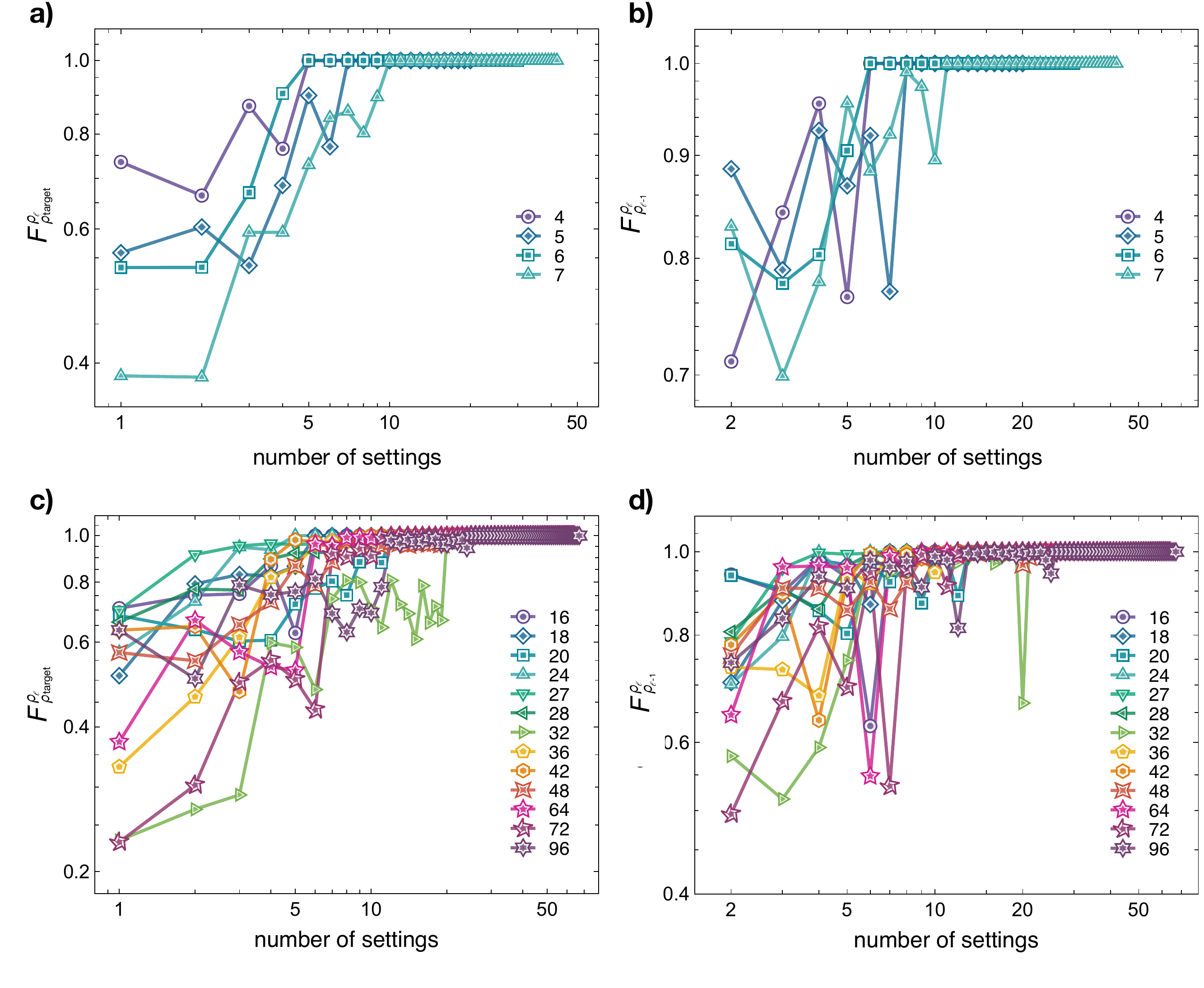}
	\caption{\label{fig:ECT_fig1.5_hres_notQST} {\bf a)} Reconstruction of a noiseless $N$-qubit $W$ state with $N={4,5,6,7}$.  \(F^{\rho_{\ell}}_{\rho_{\mathrm{target}}} \) represents the fidelity with respect to the target density matrix $\rho_{\mathrm{target}}$ of the ECT-QST reconstructed $\rho_{\ell}$ as the identified settings $s\in S'_t$ are progressively measured. Notice that an extremely good reconstruction (\(F^{\rho_{\ell}}_{\rho_{\mathrm{target}}}>99.9\% \))  is obtained already for $\ell_N={5,7,5,10}$ settings, {\it i.e.}, well before exhausting all the $s\in S'_t$ . {\bf b)} Fidelity $F^{\rho_\ell}_{\rho_{\ell-1}}$ between ECT-QST reconstructions with $\ell$ and $\ell -1$ settings. {\bf c) , d)} Same as in {\bf a), b)} but for noiseless 7-qubit depth-3 random circuits producing an output state with a  number of non-zero diagonal elements $\left\{ \rho_{ii} \right\}$ between 16 and 96. On average a fidelity \(F^{\rho_{\ell}}_{\rho_{\mathrm{target}}}>99.9\% \) can be achieved considering only $\ell_7=15(7)$ settings, indicating that low-weight settings minimally affect the maximum likelihood reconstruction.}
\end{figure*}

\subsection{\label{sec:staterec}State reconstruction} 
To estimate the density matrix \( \rho \) after measuring the selected settings $s \in S'_t$, we employ a Maximum Likelihood approach. This involves finding the density matrix \( \rho \) that minimizes the likelihood function:
\begin{equation}
    L(\rho) = 
    \sum_{s,n}
    \left(
    \frac{{\cal N}^{(s)}_{n} - N^{(s)}_{n}}{2 \sqrt{{\cal N}^{(s)}_{n}}}
    \right)^2,
    \label{eq:maxlik}
\end{equation}
where \( {\cal N}^{(s)}_{n} = \langle \phi^{(s)}_n | \rho | \phi^{(s)}_n \rangle \) represents the expected measurement outcome, and \( N^{(s)}_{n} \) are the actual measured results. In Eq.~(\ref{eq:maxlik}), the index \( n \) takes \( d^N \) values for \( N \) qudits, which leads to an exponential increase in the number of terms contributing to the likelihood.

To minimize this function, a model for the density matrix \( \rho \) is required. The most general and unbiased model, based on the Cholesky decomposition, assumes \( \rho = L L^\dagger \), where \( L \) is a lower-triangular matrix of size \( {d^N} \times {d^N} \). However, as \( N \) increases, evaluating \( L(\rho) \) (and its gradient for minimization) becomes computationally demanding.

Since full-rank density matrices are typically not of interest, we found it more efficient to use \( \rho = M M^\dagger \), where \( M \) is a \( 2^N \times r \) matrix, and the rank \( r \) is kept relatively small. In our case, we fixed \( r = N \). Once the density matrix \( \rho \) is reconstructed, the validity of this approximation is checked by ensuring that the number of relevant eigenstates of \( \rho \) is smaller than \( r \). Since the inverse of the purity provides an estimate of the matrix rank, we verify whether the condition
    \( r >1/{\mathrm{tr}(\rho^2)} \),
is satisfied; if not, we increase $r$ and repeat the minimization process.

\subsection{\label{ECTvstQST}Connection with the tQST approach}

The ECT-QST approach can be viewed as an extension of the original tQST approach~\cite{tQST}, tailored for multiplexing platforms where one performs projective measurements on all the basis vectors of a given setting. This is the case, for example, with the IBMQ platform, which will be used in the following.

The connection between the tQST and ECT algorithms becomes clear when considering the rank-3 tensor $A^{(s)}_{mn}$ defined in Eq.~(\ref{eq:A}), which represents the overlap between the $n$-th basis vector of the setting $s$, $|\phi^{(s)}_n\rangle$,  and the real or imaginary part of the matrix element $(i,j)$ (denoted by the multi index $m$).
Since the setting $s$ is related to the matrix element $m$ according to the procedure described in {\it ii)} above, the projector $|\phi^{(s)}_n\rangle$ providing the most information about the matrix element $m$ is the one with the maximum overlap, {\it i.e.}, the index $n_\mathrm{max}$ given by:
\begin{align}
n_\mathrm{max} &= \mathrm{argmax}\left( | A^{(s)}_{mn} |^2 \right).
\label{eq:nmax}
\end{align}

In this way, the original tQST approach is recovered as a two-step process: first, the setting $s$ that provides the most information about any matrix element $m$ of interest is identified, and the matrix from Eq.~(\ref{eq:A}) is constructed. Then, the most relevant vector is determined by Eq.~(\ref{eq:nmax}). 
In general, there may not be a unique solution to Eq.~(\ref{eq:nmax}), meaning that multiple vectors $|\phi_n^{(s)}\rangle$ may have the same maximum overlap. In the original tQST paper, which focused on qubits, heuristic guidelines were provided to narrow the search and achieve a unique solution; herein, for any $d$, we select the smallest value of $n$ that satisfies Eq.~(\ref{eq:nmax}).

These points clarify the relationship between tQST and ECT tomography. While tQST focuses on minimizing the number of projective measurements, ECT-QST reduces the number of measurement settings required. tQST is effective for platforms where projective measurements must be taken independently, whereas ECT-QST is optimized for multiplexing architectures, where reducing the number of settings is critical.

It is clear that the number of projective measurements used in ECT-QST is ${\cal O}(d^N)$ times larger than in tQST, introducing a significant amount of redundancy. However, as we will see, this redundancy makes ECT-QST more robust in noisy environments compared to tQST.

\section{Results}

We now demonstrate the efficiency of the ECT-QST approach. 

\subsection{Noiseless case}

We first examine the case in which there is no noise in the preparation of the target state; 
In this case, a suitable choice for the threshold $t$ is the value of the smallest non-zero element of the diagonal of the target state density matrix. In view of the implementation of the algorithm on the IBMQ platform, we will only consider $N$-qubit systems ($d=2$) with $4\leq N\leq7$.

\noindent{\it i) GHZ states} -- Regardless of $N$, a $d$-qudit GHZ state has $d(d-1)/2$ off-diagonal elements, resulting in $d(d-1)\sim{\cal O}(d^2)$ settings to be reconstructed within the ECT-QST algorithm. In the qubit case at hand, only $3\sim{\cal O}(1)$ settings are necessary. Besides the initial setting $s^{(00\dots0)}$ that measures the diagonal elements,  \( S'_t \) includes only two more settings: one that measures the real part of the element at \( i=0=\underbrace{00\cdots0^{2}}_{N \text{ times}} \) and \( j=2^N-1=\underbrace{11\cdots1^{2}}_{N \text{ times}} \), \( s^{(11\cdots1)}=\sigma_{x}\otimes\sigma_{x}\otimes\cdots\otimes\sigma_{x} \);  and the corresponding \( s^{(21\cdots1)}=\sigma_{y}\otimes\sigma_{x}\otimes\cdots\otimes\sigma_{x} \) for its imaginary~part (which, in fact, would not be needed as the state is purely real). In the absence of noise, the ECT-QST algorithm reconstructs the analyzed GHZ states with a perfect 100\% fidelity.\\
\noindent{\it ii) $W$ States} --- Independently from the system dimensionality $d$, the ECT-QST procedure identifies $N(N-1)\sim{\cal O}(N^2)$ settings to reconstruct an $N$-qudit $W$ state. In the qubit case at hand, we show in Fig.~\ref{fig:ECT_fig1.5_hres_notQST} a) and b),  the fidelity \(F^{\rho_{\ell}}_{\rho_{\mathrm{target}}} \) between the ECT-QST reconstruction $\rho_{\ell}$ with $\ell$ settings and the target density matrix \(\rho_{\mathrm{target}}\), and the fidelity \(F^{\rho_{\ell}}_{\rho_{\ell-1}}\) between ECT-QST reconstructions with $\ell$ and $\ell -1$ settings -- from the ordered list described in Sec.~\ref{sec:sorting} --, respectively. An extremely good reconstruction (\(F^{\rho_{\ell}}_{\rho_{\mathrm{target}}}>99.9\% \))  is obtained already for $\ell_N={5,7,5,10}$ settings (for $N={4,5,6,7}$ respectively), {\it i.e.}, well before exhausting all the identified settings  $s\in S'_t$.\\
\noindent{\it iii) Random states} ---  Finally, we consider pure states generated by 7-qubit depth-3 random circuits. These states have a number of non-zero diagonal elements $\left\{ \rho_{ii} \right\}$ between 16 and 96, with the cardinality $|S'_t|$ varying between 17 and 67. On average, we obtain an extremely good  (\(F^{\rho_{\ell}}_{\rho_{\mathrm{target}}}>99.9\% \)) state reconstruction with $\ell_7=15(7)$; this fast saturation of the fidelity is driven by the sorting of the settings $s$ according to their weights $w_{m}(s)$, indicating that, as expected, low-weight settings minimally affect the maximum likelihood reconstruction.

\subsection{IBMQ implementation.}

The same states analyzed in the previous subsections have been implemented on the IBMQ superconducting platform. To account for the noise level of the platform and allow for a direct comparison with the tQST results reported in~\cite{tQST}, the threshold $t$ has been chosen according to the algorithm described in~\cite{tQST}, and reported for the reader's convenience in Appendix~\ref{app:threshold-IBMQ}. All the results have been obtained using $n=10^4$ shots.

\begin{figure}[!t]
	\centering
	\includegraphics[scale=0.265]{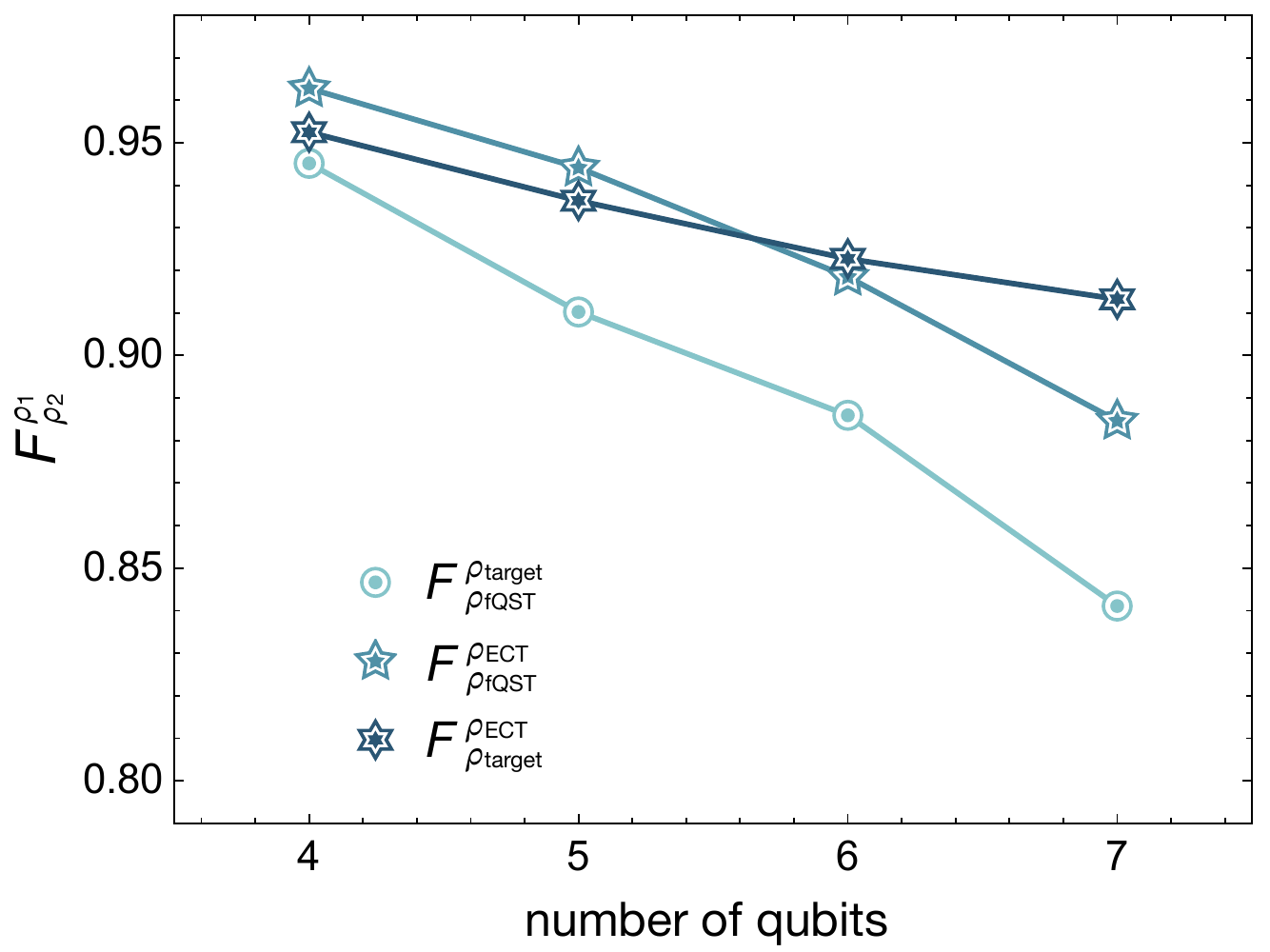}
	\caption{\label{fig:ECT_fig1_hres_notQST} Fidelities of the reconstructed density matrix from the output of an IBMQ circuit targeting the generation of a $N$-qubit GHZ state with $N={4,5,6,7}$.
	}
\end{figure}

\begin{figure*}[!t]
	\centering
	\includegraphics[scale=0.265]{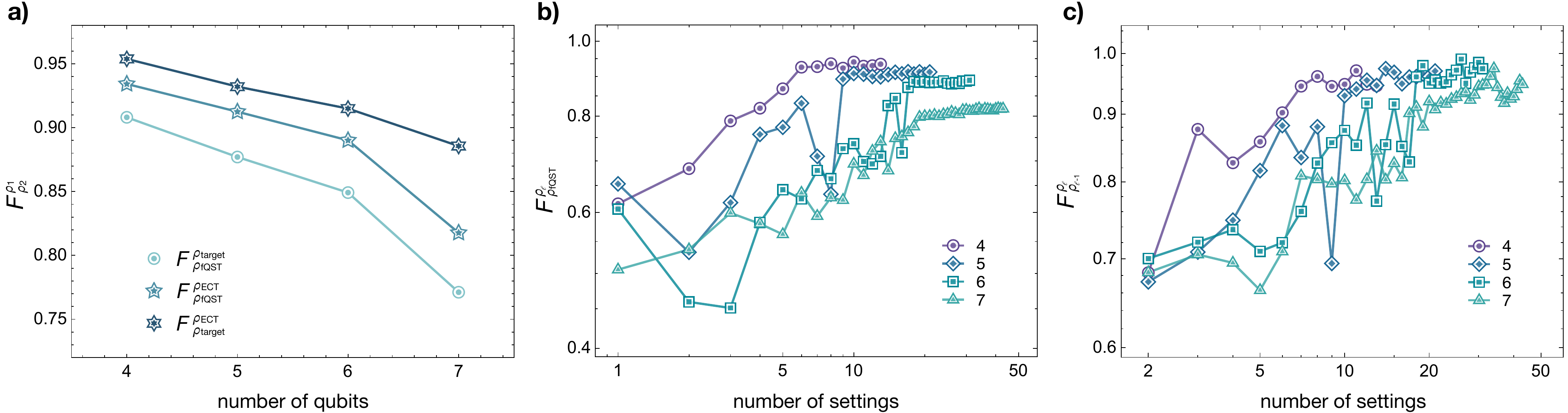}
	\caption{\label{fig:ECT_fig2_hres_notQST} {\bf a)} Same as in Fig.~\ref{fig:ECT_fig1_hres_notQST} but for the output of an IBMQ circuit targeting the generation of a  $N$-qubit $W$ state with $N={4,5,6,7}$.  Degrading of the fidelity as $N$ increases, is due to  relaxation processes that populate the unwanted ground-state component (see the discussion in Appendix~\ref{app:W}, and, in particular, Fig.~\ref{fig:ECT_SM_Fig2}). {\bf b)} Fidelity with respect to the fQST density matrix of the ECT-QST reconstructed $\rho_{\ell}$ as the identified settings $s\in S'_t$ are progressively measured. Settings barely affecting the fidelity coincide with the ones measuring imaginary parts. {\bf c)} Fidelity $F^{\rho_\ell}_{\rho_{\ell-1}}$ between ECT-QST reconstructions with $\ell$ and $\ell -1$ settings.}
\end{figure*}

\noindent{\it i) GHZ states} --- To generate an $N$-qubit GHZ state, we used the circuit described in~\cite{Cruz_2019}. This circuit has logarithmic time complexity, minimizing the required depth. Results are shown in Fig.~\ref{fig:ECT_fig1_hres_notQST}, where we plot the fidelity $F^{\rho_{\mathrm{ECT}}}_{\rho_{\mathrm{fQST}}}$ between the ECT-QST reconstructed density matrix, $\rho_{\mathrm{ECT}}$, and the full QST reconstructed one, $\rho_{\mathrm{fQST}}$. 
As can be seen, the ECT-QST algorithm reconstructs the density matrix of GHZ states with a fidelity bigger than 90\% in all cases.

\noindent{\it ii) W States} --- 
To generate $N$-qubit $W$ states we leveraged the circuit with logarithmic time complexity proposed in Ref.~\cite{Cruz_2019}  (see, however, the discussion in Appendix~\ref{app:W}).
In Fig.~\ref{fig:ECT_fig2_hres_notQST}a), we show the fidelity  $F^{\rho_{\mathrm{ECT}}}_{\rho_{\mathrm{fQST}}}$ as was done in Fig.~\ref{fig:ECT_fig1_hres_notQST}; clearly , we achieve  lower fidelities than those obtained int the GHZ case. This is due to the fact that although the implemented circuits generate reasonably trusty $W$ states on the IBMQ platform, the population of the $\ket{0}^{\otimes N}$ component ({\it i.e.}, the ground state) increases significantly as $N$ grows (Appendix~\ref{app:W}, Fig. 5). This additional unwanted component, motivates in turn the fidelity loss observed in Fig.~\ref{fig:ECT_fig2_hres_notQST}a).

Panels b) and c) in Fig.~\ref{fig:ECT_fig2_hres_notQST}   show, respectively, the fidelity \(F^{\rho_{\ell}}_{\rho_{\mathrm{fQST}}} \) between the ECT-QST reconstruction $\rho_{\ell}$ with $\ell$ settings and the full QST reconstruction \(\rho_{\mathrm{fQST}}\), and the fidelity \(F^{\rho_{\ell}}_{\rho_{\ell-1}}\) between ECT-QST reconstructions with $\ell$ and $\ell -1$ settings. Notice that settings that, when measured, do not significantly increase the fidelity are related to imaginary parts of the density matrix elements (recall that these states are real). 

\noindent{\it iii) Random states} --- 
Finally, we implemented on the IBMQ the same pure states generated by 7-qubit depth-3 random circuits  studied in the noiseless case. 

\begin{figure*}[!t]
	\centering
	\includegraphics[scale=0.265]{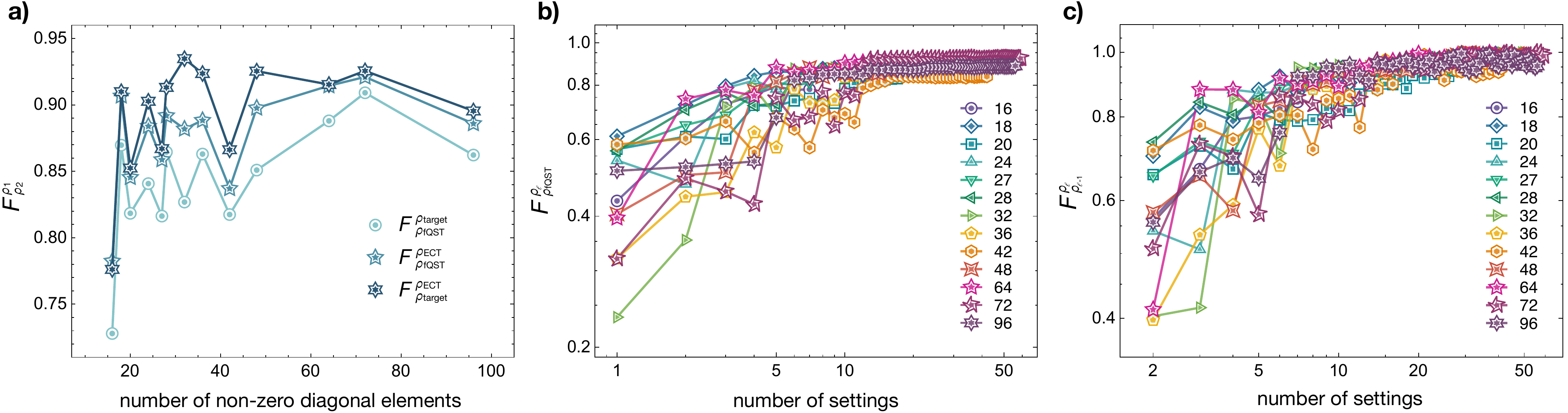}
	\caption{\label{fig:ECT_fig3_hres_notQST} {\bf a)} Fidelities of the reconstructed density matrix of $7$-qubit IBMQ random circuits of depth-3 producing an output state with a variable number of non-zero diagonal elements $\left\{ \rho_{ii} \right\}$. {\bf b)} and {\bf c)}: Same as in Fig.~\ref{fig:ECT_fig2_hres_notQST} for the same 7-qubit circuits. 
	}
\end{figure*}

Results are shown in Fig.~\ref{fig:ECT_fig3_hres_notQST}, 
with a full comparison between fQST, tQST and ECT-QST reconstructions for this case and  the additional cases $N=4,5$ and $6$ is reported  in Appendix~\ref{app:fullres}.
We found that ECT-QST reduces the number of required settings by a factor of 10 to 100 compared to fQST, and by a factor of 1 to 20 compared to the original tQST algorithm.

\begin{table}[!t]
    \begin{center}
    \begin{tabular}{
        >{\centering\arraybackslash}m{4mm}|| 
        >{\raggedleft\arraybackslash}m{9mm}| 
        >{\raggedleft\arraybackslash}m{10mm}| 
        >{\raggedleft\arraybackslash}m{10mm}|| 
        >{\raggedleft\arraybackslash}m{9mm}| 
        >{\raggedleft\arraybackslash}m{10mm}| 
        >{\raggedleft\arraybackslash}m{10mm}|| 
        >{\raggedleft\arraybackslash}
        	m{7.5mm} 
    }
    $N$ & \centering$|S_t'|$ & \centering$\overline{F^{\rho_\mathrm{ECT}}_{\rho_{\mathrm{fQST}}}}$ & \centering$\overline{F^{\rho_\mathrm{ECT}}_{\rho_{\mathrm{target}}}}$ & \centering$|S'_{t,\ell^*}|$ & 
    	\centering$\overline{F^{\rho_{\ell^*}}_{\rho_{\mathrm{fQST}}}}$ & 
    	\centering$\overline{F^{\rho_{\ell^*}}_{\rho_{\mathrm{target}}}}$ &$\sim rN$ \\
        \hline\hline
        4 & 13(5) & 97(1)\% & 96(1)\% &  7(3) & 96(2)\% & 95(3)\% & 5 \\
        5 & 25(5) & 95(1)\% & 95(1)\% & 10(3) & 94(1)\% & 94(2)\% & 6 \\
        6 & 35(7) & 92(3)\% & 94(2)\% & 12(5) & 91(3)\% & 93(3)\% & 8 \\
        7 & 40(14)& 87(4)\% & 89(5)\% & 18(6) & 87(2)\% & 89(4)\% & 11
    \end{tabular}
    \caption{\label{tab:comparison}
    Performance of ECT-QST in the case of quantum states corresponding to $N=4,5,6,7$ depth-3 random circuits.
    $|S_t'|$ is the number of settings determined by the pruning procedure, $\overline{F^{\rho_\mathrm{ECT}}_{\rho_{\mathrm{fQST}}}}$ is the average fidelity between the ECT and fQST reconstructed states, and $\overline{F^{\rho_\mathrm{ECT}}_{\rho_{\mathrm{target}}}}$ is the average fidelity of the ECT reconstructed state with the (ideal) target state.
    $|S'_{t,\ell^*}|$ is the number of settings used when progressively measuring the settings in $S_t'$ ordered by their decreasing weight and terminating when $\overline{F^{\rho_{\ell^*}}_{\rho_{\ell^*-1}}} > 95\%$.
    $\overline{F^{\rho_{\ell^*}}_{\rho_{\mathrm{fQST}}}}$ and $\overline{F^{\rho_{\ell^*}}_{\rho_{\mathrm{target}}}}$ are the fidelities with respect to fQST and the (ideal) target state obtained at the end of this procedure, respectively.
    The last column report an estimate of $rN$ where $r$ is calculated as the average of the inverse purity of the fQST reconstructed density matrices; this value corresponds to the number of settings expected by the Adaptive Compressed Sensing approach of Refs.~\onlinecite{ahn2019adaptive, ahn2019adaptive2}.
    Number in parentheses denote the standard uncertainty in the last digit(s).}
    \end{center}
\end{table}

\section{Discussion and conclusions}

The ECT-QST algorithm enables the efficient reconstruction of a quantum state's density matrix while minimizing the number of required measurement settings. $d$-qudit GHZ states represent the most favorable scenario, requiring only \({\cal O}(d^2)\) settings (independently of the system size $N$) for a complete reconstruction. This corresponds to ${\cal O}(1)$ in the qubit case: to the best of our knowledge, no current method surpasses this efficiency. 

On the other hand, despite the sparsity of the density matrix for $W$ states, ECT-QST still requires \({\cal O}(N^2)\) settings (independently of $d$, and similarly to the original tQST algorithm), as each off-diagonal element requires two settings.

In other cases, ECT-QST can require as few as \({\cal O}(N)\) settings. For example, in the case of the 7-qubit color-code state studied in Ref.~\cite{Bombin:2006sj}, only 15 settings are needed for a faithful reconstruction of the $\vert\overline{0}\rangle$ and $\vert\overline{1}\rangle$ states, compared to the 57 required by tQST~\cite{Maragnano23} and the 127 used in Ref.~\cite{riofrio2017experimental}.

Finally, for random circuits, the number of settings heavily depends on the number of non-zero diagonal elements. 
 
In all cases analyzed, the sorting of settings according to their weights results in a rapid convergence of the fidelity \(F^{\rho_\ell}_{\rho_{\ell-1}}\). This allows for (heuristic) strategies to further reduce the required number of settings for reconstructing a given state. For example, one can estimate the density matrix \( \rho_\ell \) at each step \(\ell = 1, 2, \ldots\), continuing to add settings until the fidelity \( F^{\rho_\ell}_{\rho_{\ell-1}} \) stabilizes and reaches a predefined value \(F^*\) at a specific number of settings \(\ell^*\), and remains stable when adding more settings. For instance, selecting \(F^* = 95\%\) and requiring stability when adding three additional settings yields the results shown in Table~\ref{tab:comparison}.  Evidently, ECT-QST results in a linear scaling with the number of qubits, similar to adaptive compressed sensing methods, but without the need for basis changes between measurements, which can be experimentally challenging.

In summary, ECT-QST is an enhanced version of the tQST approach, designed to minimize the number of settings required for faithful reconstruction of the density matrix for an $N$-qudit system. 
Its key innovations include: 1) associating two measurement settings per matrix element (one for the real part and one for the imaginary part), using only standard 1-qudit observables (Pauli operators for qubits); 2) reducing the number of settings to the minimum necessary for complete reconstruction; and 3) employing a sorting algorithm that progressively improves the accuracy of the density matrix as more settings are measured.

Implementing of the ECT-QST algorithm in photonic platforms is currently under way.

\acknowledgments
Work supported by PNRR MUR project PE0000023-NQSTI. We thank D.~Cruz for useful discussions in connection to the algorithm to generate $W$-states described in~\cite{Cruz_2019}.

\appendix

\section{Examples of ECT-QST}
Let us illustrate how ECT-QST would perform the tomography of the 2-qutrit state
\begin{align}
    |\Psi\rangle &= \frac{1}{\sqrt{2}} |00\rangle +
    \frac{1}{\sqrt{3}}  |02\rangle +
    \frac{1}{\sqrt{12}} |11\rangle +
    \frac{i}{\sqrt{12}} |12\rangle,
\end{align}
assuming perfect measurements. (The 1-qutrit operators are the SU(3) matrices provided in Sect.~\ref{sec:1-qudit}). The first step, that is the measurement of the diagonal of the density matrix using the setting $s^{(00)}$ would result into 4 elements out of nine being different from zero. When we represent $|\Psi\rangle$ in $\mathbb{C}^9$, these elements are those corresponding to the indices $0$ (with value $|\langle 00|\Psi\rangle|^2$), $2$ (with value $|\langle 02|\Psi\rangle|^2$), $4$ (with value $|\langle 11|\Psi\rangle|^2$), and $5$ (with value $|\langle12|\Psi\rangle|^2$). 
Any threshold $t < 1/12$ would therefore select 6 matrix elements for further consideration, namely $(i,j) = (0,2),\ (0,4),\ (0,5),\ (2,4),\ (2,5),$ and $(4,5)$ with 12 associated settings: 6 for the real and 6 for the imaginary part of each matrix element

The setting corresponding to the matrix element $(i,j)=(0,2)$ can be identified considering the base-3 representation $i^3=00^3$ and $j^3=02^3$, and using Eq. (\ref{eq:S_base_d}) of the main text. One would have to consider the setting measuring $\sigma^{(0)}$ for the first qutrit and the $\sigma^{(2)}$ for the second qutrit -- represented in our notation by $s^{(02)}$ -- to gain information on the real part, and the corresponding setting $s^{(05)}$ for the imaginary part.
Analogously, the settings providing information on the matrix element $(0,4)$ are $s^{(11)}$ for the real part and $s^{(41)}$ for the imaginary part. Proceeding in this way one finds 12 settings: $s^{(02)},\ s^{(11)},\ s^{(12)},\ s^{(13)},\ s^{(10)},\ s^{(03)}$ giving information on the real part of the density matrix and $s^{(05)},\ s^{(41)},\ s^{(42)},\ s^{(43)},\ s^{(40)},\ s^{(06)}$ for the imaginary part.

The corresponding matrix $C_{sm}$ is
\begin{equation}
C_{sm} = \left(
\begin{array}{cccccccccccc}
\frac{1}{2} & 0 & 0 & 0 & 0 & 0 & 0 & 0 & 0 & 0 & 0 & 0 \\
0 & \frac{1}{4} & 0 & 0 & \frac{1}{2} & 0 & 0 & 0 & 0 & 0 & 0 & 0 \\
\frac{1}{4} & 0 & \frac{1}{4} & 0 & \frac{1}{4} & 0 & 0 & 0 & 0 & 0 & 0 & 0 \\
0 & 0 & 0 & \frac{1}{4} & \frac{1}{4} & \frac{1}{4} & 0 & 0 & 0 & 0 & 0 & 0 \\
0 & 0 & 0 & 0 & \frac{1}{2} & 0 & 0 & 0 & 0 & 0 & 0 & 0 \\
0 & 0 & 0 & 0 & 0 & \frac{1}{2} & 0 & 0 & 0 & 0 & 0 & 0 \\
0 & 0 & 0 & 0 & 0 & 0 & \frac{1}{2} & 0 & 0 & 0 & 0 & 0 \\
0 & 0 & 0 & 0 & 0 & 0 & 0 & \frac{1}{4} & 0 & 0 & \frac{1}{2} & 0 \\
\frac{1}{4} & 0 & 0 & 0 & 0 & 0 & 0 & 0 & \frac{1}{4} & 0 & \frac{1}{4} & 0 \\
0 & 0 & 0 & 0 & 0 & \frac{1}{4} & 0 & 0 & 0 & \frac{1}{4} & \frac{1}{4} & 0 \\
0 & 0 & 0 & 0 & 0 & 0 & 0 & 0 & 0 & 0 & \frac{1}{2} & 0 \\
0 & 0 & 0 & 0 & 0 & 0 & 0 & 0 & 0 & 0 & 0 & \frac{1}{2}
\end{array}
\right)
\label{eq:C1}
\end{equation}
where the index $s$ such that settings corresponding to the real part come first, followed by those corresponding to the imaginary part; and, similarly, the matrix elements $m$, which these settings provide information on, are arranged with the real part first, followed by the imaginary part. The target vector $\beta_m$ for the pruning procedure is
\begin{equation}
\beta_m = \mathrm{max}_{s}\,C_{sm} = \left(
\begin{array}{cccccccccccc}
\frac{1}{2} & \frac{1}{4} & \frac{1}{4} & \frac{1}{4} & \frac{1}{2} & \frac{1}{2} & \frac{1}{2} & \frac{1}{4} & \frac{1}{4} & \frac{1}{4} & \frac{1}{2} & \frac{1}{2}
\end{array}
\right).
\end{equation}
The pruning procedure described in the main text indicates that the settings $s^{(10)}$  and $s^{(40)}$ (corresponding to the fifth and eleventh row of the matrix $C$) can be removed.
In fact, one can verify that the sum of all the rows of the matrix $C$ in Eq.~(\ref{eq:C1}) except row 5 and 11 is still a vector whose elements are greater than or equal to the corresponding elements of the vector $\beta$.
In this case, the ECT-QST approach indicates that 10 additional settings are required to perform the tomography of the state $|\Psi\rangle$, once the diagonal of the density matrix has been measured.

Similarly, let us consider how ECT-QST would proceed in the case of
\begin{align}
    |\Phi\rangle &= 
    \frac{1}{\sqrt{2}} |00\rangle +
    \frac{1}{\sqrt{3}}  |02\rangle +
    \frac{1}{\sqrt{12}} |10\rangle +
    \frac{i}{\sqrt{12}} |12\rangle.
\end{align}
In this case, the $\mathbb{C}^9$ representation has non-zero entries only at indices $0,\ 2,\ 3$, and $5$. Working out the corresponding settings according to Eq. (\ref{eq:S_base_d}) of the main text, one find only 6 independent settings: $s^{(02)}$ ($s^{(05)}$) measuring the real (imaginary) parts of the (0,2) and (3,5) density matrix elements; $s^{(10)}$ ($s^{(40)}$), for the real (imaginary) parts of elements (0,3) and (2,5); and finally, $s^{(12)}$ ($s^{(42)}$) for the real (imaginary) parts of  elements (0,5) and (2,3).
In this case the matrix $C_{sm}$ is then
\begin{equation}
    C_{sm} = \left(
    \begin{array}{cccccccccccc}
\frac{1}{2} & 0 & 0 & 0 & 0 & \frac{1}{2} & 0 & 0 & 0 & 0 & 0 & 0 \\
0 & \frac{1}{2} & 0 & 0 & \frac{1}{2} & 0 & 0 & 0 & 0 & 0 & 0 & 0 \\
\frac{1}{4} & \frac{1}{4} & \frac{1}{4} & \frac{1}{4} & \frac{1}{4} & \frac{1}{4} & 0 & 0 & 0 & 0 & 0 & 0 \\
0 & 0 & 0 & 0 & 0 & 0 & \frac{1}{2} & 0 & 0 & 0 & 0 & \frac{1}{2} \\
0 & 0 & 0 & 0 & 0 & 0 & 0 & \frac{1}{2} & 0 & 0 & \frac{1}{2} & 0 \\
\frac{1}{4} & 0 & 0 & 0 & 0 & \frac{1}{4} & 0 & \frac{1}{4} & \frac{1}{4} & \frac{1}{4} & \frac{1}{4} & 0 \\
\end{array}
\right),
\end{equation}
which has a structure that prevents further pruning. In this case, the way we associate one setting to each matrix element above threshold already produces a number of settings smaller than the number of matrix elements to be determined.

\section{\label{app:W}Generation of W states} 
A logarithmic time complexity circuit for generating $N$-qubit $W$ states, has been proposed in~\cite{Cruz_2019}. This circuit minimizes depth, and is based on a fundamental two-qubit gate block $B(p)$ $(0 < p < 1)$:
\begin{figure}[!h]
	\includegraphics[scale=0.4]{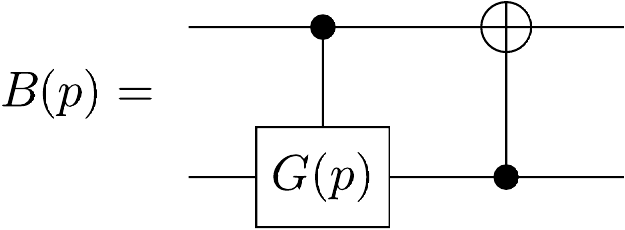}
\end{figure}

\noindent The block consists of a controlled-$G(p)$ rotation (equivalent to a controlled rotation \( U (2 \arccos\sqrt{p}, 0, 0, 0) \) on an IBMQ processor) followed by an inverted CNOT. The gate $B(p)$ operates as follows:
\begin{subequations}
\begin{align}
B(p)\ket{00}& =\ket{00};\\
B(p)\ket{10}& =\sqrt{p} \ket{10}+\sqrt{1-p} \ket{01}.
\end{align}
\end{subequations}
An algorithm to construct the circuit for a $N$-qubit system, is then described in the Appendix of the same paper. However, it should be noticed that this algorithm led to incorrect states for certain numbers of qubits; for example, up to 20-qubit systems, one generates the wrong density matrix for $N=$10, 14, 18, 19 and 20. With reference to the original text, the correct algorithm can be obtained by: 
\begin{itemize}
	\item Removing line 4;
	\item Replace line 5 with: ``Each leaf $(n,m)$ generates an upper child $(f(n/2),n)$ and a lower child $(f(m/2-n/2),m-n)$, where the function $f$ returns the lower integer part of its argument'';
	\item Removing lines 8 and 9 (as swapping is no longer necessary). 
\end{itemize} 

\begin{figure}[!t]
	\includegraphics[scale=0.45]{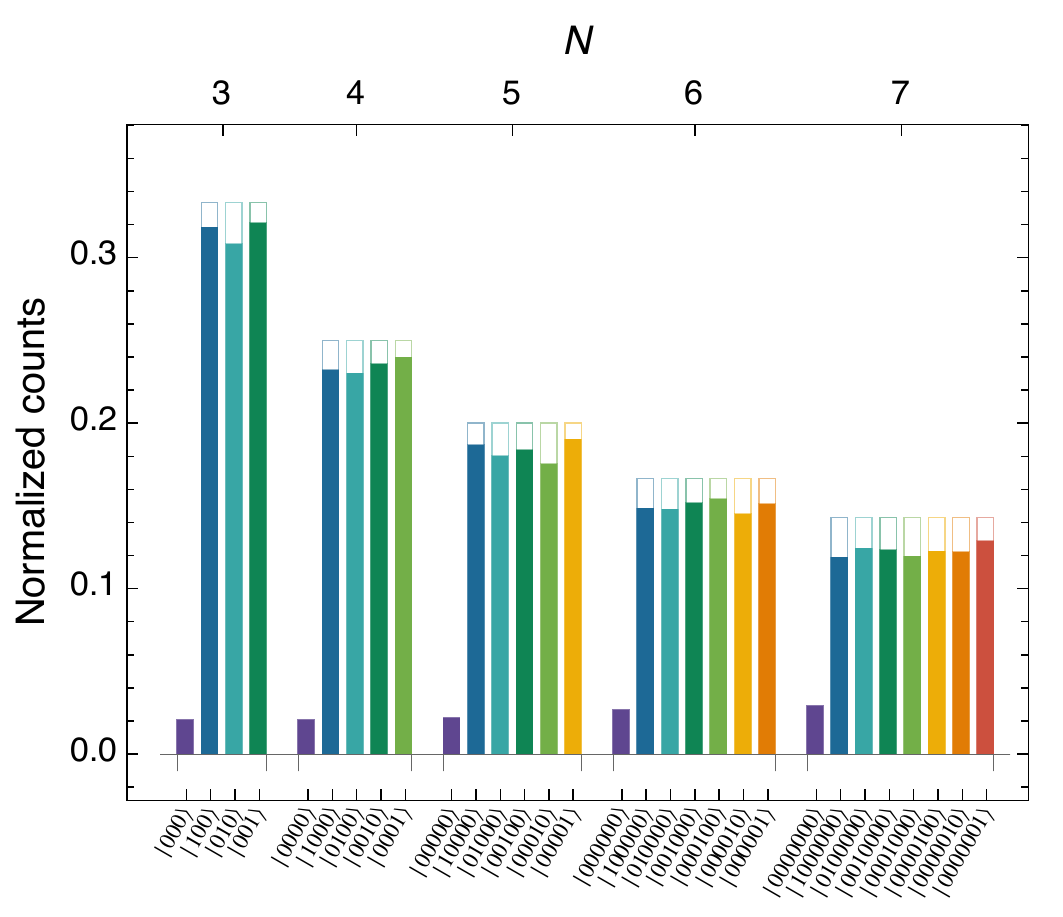}
	\caption{\label{fig:ECT_SM_Fig2} Normalized counts for the expected non-zero density matrix elements plus the ground-state of a $W$-state generated with the corrected   version of the logarithmic time complexity circuit of~\cite{Cruz_2019}. Relaxation effects drive the ground state non-zero population, whose relative importance becomes bigger as $N$ increases.}
\end{figure}

As can be appreciated from Fig.~\ref{fig:ECT_SM_Fig2}, the generated state shows a non zero population for the ground state. This effect can be attributed to relaxation processes favouring the ground state: the longer the state preparation, the more pronounced the relaxation~\cite{Cruz_2019}. This has two effects on our reconstructions:
\begin{enumerate}
	\item On the one hand, as for $N \leq 7$, the $\ket{0}^{\otimes N}$ component remains below threshold, ECT-QST reconstructs a density matrix closer to the target state (where \(\mbox{}^{N\otimes}\hspace{-0.1cm}\bra{0}\rho_{\mathrm{ex}}\ket{0}\hspace{-0.07cm}\mbox{}^{\otimes N} = 0\)) than the fQST one (where \(\mbox{}^{N\otimes}\hspace{-0.1cm}\bra{0}\rho_{\mathrm{FT}}\ket{0}\hspace{-0.07cm}\mbox{}^{\otimes N} \neq 0\), with a sizeable ground state population for large $N$);
	\item On the other hand, the fidelity with respect to the exact density matrix \(\rho_{\mathrm{ex}}\) degrades as $N$ increases, due to the shift in population toward the ground state.
\end{enumerate}

\section{\label{app:threshold}Choice of threshold.}

The selection of an appropriate threshold value is dependent on the specific physical system used to implement the qubits ({\em e.g.}, noise level), the amount of available resources ({\em e.g.}, time requirements), and the desired quantum state to be generated.

\subsection{\label{app:threshold-IBMQ} IBMQ platform}

For the reader's convenience we reproduce here the algorithm identified in~\cite{tQST} for selecting an appropriate circuit-specific threshold on the IBMQ platform.

Using the IBMQ simulator available in the {\tt qiskit} package, one first simulates the unitary evolution of a ground-state initialized quantum register according to the circuit itself (we have used $n=10^4$ shots). Measuring all qubits yields the expected diagonal counts in the absence of errors, which can be separated into zero and non-zero counts. Second, one uses the IBMQ simulator (which includes the effect of noise) to run the circuit a number of times (100 in our case) and record: the maximum value of the counts among the expected zero elements of the diagonal, $c_0^{\mathrm{max}}$; and the minimum value of the counts for the smallest expected non-zero diagonal element, $c_{>0}^{\mathrm{min}}$. Third, one defines, in a conservative way: the {\it noise threshold} as $t_0 = c_0^{\mathrm{max}} + N\sqrt{c_0^{\mathrm{max}}}$; and the {\it signal threshold} as $t_{>0} = c_{>0}^{\mathrm{min}} - N\sqrt{c_{>0}^{\mathrm{min}}}$. The square root terms consider the variability of the counts $c_0^{\mathrm{max}}$, $c_{>0}^{\mathrm{min}}$ each time the circuit is simulated; the $N$ factor takes finally into account that, for the quantum processors considered, the noise increases with the number of qubits $N$. Then, we use as the circuit-specific (normalized) threshold the quantity
\begin{align}
t = \mathrm{max}(t_0, t_{>0})/n,
\end{align}
which discards those diagonal entries most affected by noise.

\subsection{\label{app:threshold-Gini}Gini index}

In the absence of a method to estimate the optimal value of the threshold $t$ for a given measurement of the density matrix's diagonal elements, analyzing the sparsity of the latter can provide useful guidance. In \cite{hurley2009comparingmeasuressparsity}, various measures of sparsity are examined; among these, the Gini index can be identified as the measure that meets the criteria necessary for establishing a threshold.

Let \( \mathbf{c} \) be a vector with components \( c_i,\ i=1,\dots,n \), sorted in ascending order. The associated Gini index is defined as
\begin{align}
	{\mathrm{GI}}(\mathbf{c}) = 1 - 2 \sum_{k=1}^{n} \frac{c_{k}}{\| \mathbf{c} \|_1} \left( \frac{n - k + 1}{2 n} \right),
\end{align}
where \( \| \mathbf{c} \|_p \) represents the \( p \)-norm, {\it i.e.}, \( \| \mathbf{c} \|_p = \left( \sum_i c_i^p \right)^{\frac{1}{p}} \) (with \(0 \leq p \leq 1\)). This index measures the inequality among the values of \( \mathbf{c} \), treated as a frequency distribution. It satisfies the bound:
\begin{align}
	0 \leq {\mathrm{GI}}(\mathbf{c}) \leq 1 - \frac{1}{n},
\end{align}
with the lower bound corresponding to a uniform vector with equal \( 1/n \) entries (perfect equality), and the upper bound corresponding to a vector with a single non-zero entry (maximal inequality).

Given a measured diagonal \(\{\rho_{ii}\}\) of the density matrix describing an \( N \)-qudit system (where \( n = d^N \)), the tQST threshold is then defined as
\begin{align}
	t = \frac{\mathrm{GI}(\{\rho_{ii}\})}{d^N - 1}.
\end{align}

We found that the threshold identified through the rescaled Gini index above when used on the IBMQ platform identifies a slightly larger number of settings (up to twice as many in the worst-case scenario) with respect to the optimal circuit-specific threshold discussed in~\ref{app:threshold-IBMQ}.

\section{\label{app:fullres}Full results for $N=4,\dots,7$ qubits random states} 

\begin{figure*}[!t]
	\centering
	\includegraphics[scale=0.265]{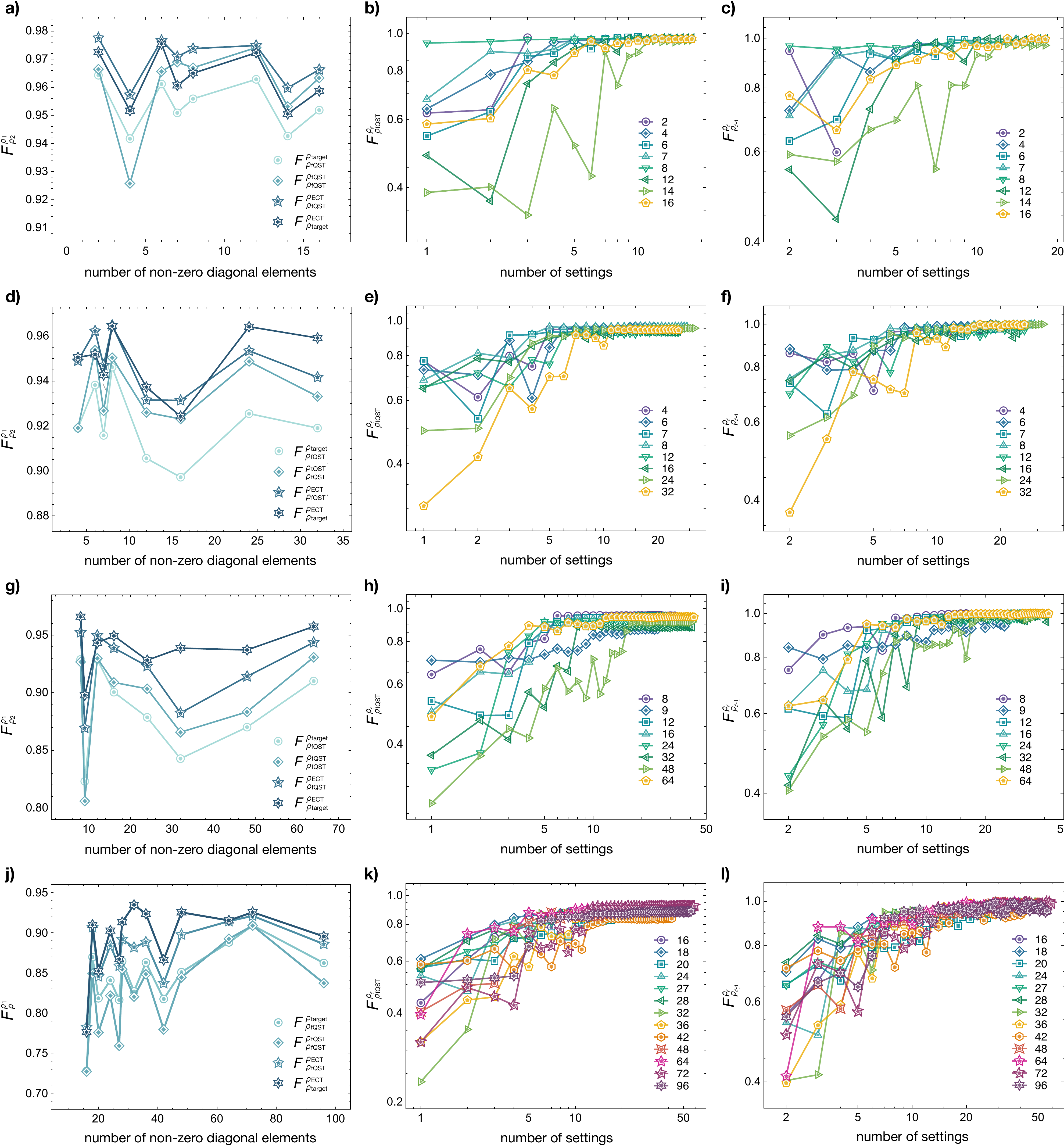}
	\caption{\label{fig:ECT_SM_Fig2_hres} {\bf a)} Fidelities of the reconstructed density matrix of $4$-qubit IBMQ random circuits of depth-3 producing an output state with a variable number of non-zero diagonal elements. {\bf b)} and {\bf c)}: Same as in Fig.~\ref{fig:ECT_fig2_hres_notQST} for the 4-qubit circuits displayed in panel {\bf a)}. Different symbols refer to the number of non-zero elements on the state diagonal (indicated in the legend). {\bf d)} -- {\bf l)}: Same as in panels {\bf a)} -- {\bf c)} for $N=5,6,7$.
	}
\end{figure*}

In Fig.~\ref{fig:ECT_SM_Fig2_hres}, we present the ECT-QST density matrix reconstruction of pure states generated by depth-3 random circuits for systems with 4 to 7 qubits, featuring varying numbers of non-zero diagonal elements. Compared to tQST, ECT-QST reproduces the fQST density matrix with slightly better fidelity (though close), while using significantly fewer settings but requiring many more measurements; see Tables~\ref{tab:tQSTres-4}--\ref{tab:tQSTres-7}.

For real states like GHZ and $W$ states, ECT-QST demonstrates greater resilience to noise. Due to the lower number of measurements required, the tQST maximum likelihood reconstruction often captures the modulus of the density matrix elements but fails to accurately reconstruct the real and imaginary parts, especially for systems with \(N > 5\). For $W$ states, this issue can be partially mitigated by significantly reducing the rank \(r\) of the initial matrix $M$. However, in the case of GHZ states, this approach proved insufficient.

\begin{widetext}
\begin{table*}[!b]
    \begin{center}
    \begin{tabular}{
        >{\centering\arraybackslash}m{10mm}| 
        >{\centering\arraybackslash}m{12mm}|| 
        >{\raggedleft\arraybackslash}m{15mm}| 
        >{\raggedleft\arraybackslash}m{10mm}| 
        >{\raggedleft\arraybackslash}m{9mm}|| 
        >{\raggedleft\arraybackslash}m{15mm}| 
        >{\raggedleft\arraybackslash}m{15mm}| 
        >{\raggedleft\arraybackslash}m{10mm}| 
        >{\raggedleft\arraybackslash}m{10mm}|| 
        >{\raggedleft\arraybackslash}m{15mm}| 
        >{\raggedleft\arraybackslash}m{15mm}| 
        >{\raggedleft\arraybackslash}m{10mm}| 
        >{\raggedleft\arraybackslash}m{10mm}  
    }
        \multicolumn{2}{c||}{} & \multicolumn{3}{c||}{\bf{fQST}} & \multicolumn{4}{c||}{\bf{tQST}} & \multicolumn{4}{c}{\bf{ECT-QST}} \\
        \cline{3-13}
        \hline
        $N$ & $\{\rho_{ii}\}\neq0$ & \centering$F^{\rho_{\mathrm{fQST}}}_{\rho_{\mathrm{target}}}$ & \centering$|S|$ & \centering$M$ & \centering$F^{\rho_{\mathrm{tQST}}}_{\rho_{\mathrm{fQST}}}$ & \centering$F^{\rho_{\mathrm{tQST}}}_{\rho_{\mathrm{target}}}$ & \centering$|S'_t|$ & \centering$M$ & \centering$F^{\rho_{\mathrm{ECT}}}_{\rho_{\mathrm{fQST}}}$ & \centering$F^{\rho_{\mathrm{ECT}}}_{\rho_{\mathrm{target}}}$ & \centering$|S'_t|$ & $M$\hspace{4mm}\mbox{} \\
        \hline\hline
        \multirow{8}{*}{4}
& 2 & 96.4\% & 81 & 1296 & 96.6\% & 96.0\% & 3 & 18 & 97.8\% & 97.3\% & 3 & 48 \\ 
& 4 & 94.2\% & 81 & 1296 & 92.6\% & 89.2\% & 9 & 28 & 95.7\% & 95.2\% & 7 & 112 \\ 
& 6 & 96.1\% & 81 & 1296 & 96.6\% & 96.3\% & 21 & 46 & 97.7\% & 97.5\% & 10 & 160 \\ 
& 7 & 95.1\% & 81 & 1296 & 96.9\% & 94.9\% & 37 & 66 & 97.1\% & 96.1\% & 14 & 224 \\ 
& 8 & 95.6\% & 81 & 1296 & 96.7\% & 96.2\% & 17 & 32 & 97.4\% & 96.5\% & 15 & 240 \\ 
& 12 & 96.3\% & 81 & 1296 & 97.4\% & 96.3\% & 79 & 194 & 97.5\% & 97.2\% & 18 & 288 \\ 
& 14 & 94.3\% & 81 & 1296 & 95.3\% & 94.2\% & 81 & 232 & 96.0\% & 95.1\% & 18 & 288 \\ 
& 16 & 95.2\% & 81 & 1296 & 96.3\% & 95.4\% & 81 & 246 & 96.6\% & 95.9\% & 18 & 288
    \end{tabular}
    \caption{\label{tab:tQSTres-4} Comparison of the fQST/tQST/ECT-QST density matrix reconstructions of output states from 4-qubit depth-3 random circuits simulated on an IBMQ system. $M = |S| \times 2^N$ is the number of projective measurements needed by each method. The circuits are characterized by different diagonal fillings, specifically the number of non-zero elements in their density matrix diagonals (second column).}
    \end{center}
\end{table*}

\begin{table*}[!b]
    \begin{center}
    \begin{tabular}{
        >{\centering\arraybackslash}m{10mm}| 
        >{\centering\arraybackslash}m{12mm}|| 
        >{\raggedleft\arraybackslash}m{15mm}| 
        >{\raggedleft\arraybackslash}m{10mm}| 
        >{\raggedleft\arraybackslash}m{9mm}|| 
        >{\raggedleft\arraybackslash}m{15mm}| 
        >{\raggedleft\arraybackslash}m{15mm}| 
        >{\raggedleft\arraybackslash}m{10mm}| 
        >{\raggedleft\arraybackslash}m{10mm}|| 
        >{\raggedleft\arraybackslash}m{15mm}| 
        >{\raggedleft\arraybackslash}m{15mm}| 
        >{\raggedleft\arraybackslash}m{10mm}| 
        >{\raggedleft\arraybackslash}m{10mm}  
    }
        \multicolumn{2}{c||}{} & \multicolumn{3}{c||}{\bf{fQST}} & \multicolumn{4}{c||}{\bf{tQST}} & \multicolumn{4}{c}{\bf{ECT-QST}} \\
        \cline{3-13}
        \hline
        $N$ & $\{\rho_{ii}\}\neq0$ & \centering$F^{\rho_{\mathrm{fQST}}}_{\rho_{\mathrm{target}}}$ & \centering$|S|$ & \centering$M$ & \centering$F^{\rho_{\mathrm{tQST}}}_{\rho_{\mathrm{fQST}}}$ & \centering$F^{\rho_{\mathrm{tQST}}}_{\rho_{\mathrm{target}}}$ & \centering$|S'_t|$ & \centering$M$ & \centering$F^{\rho_{\mathrm{ECT}}}_{\rho_{\mathrm{fQST}}}$ & \centering$F^{\rho_{\mathrm{ECT}}}_{\rho_{\mathrm{target}}}$ & \centering$|S'_t|$ & $M$\hspace{4mm}\mbox{} \\
        \hline\hline
        \multirow{8}{*}{5}
& 4 & 91.9\% & 243 & 7776 & 91.9\% & 92.1\% & 41 & 84 & 94.9\% & 95.1\% & 18 & 576 \\ 
& 6 & 93.8\% & 243 & 7776 & 95.4\% & 93.7\% & 37 & 84 & 96.2\% & 95.2\% & 19 & 608 \\ 
& 7 & 91.6\% & 243 & 7776 & 92.7\% & 92.0\% & 65 & 116 & 94.7\% & 94.3\% & 25 & 800 \\ 
& 8 & 94.6\% & 243 & 7776 & 95.0\% & 94.7\% & 115 & 212 & 96.5\% & 96.4\% & 29 & 928 \\ 
& 12 & 90.6\% & 243 & 7776 & 92.6\% & 90.3\% & 79 & 176 & 93.2\% & 93.7\% & 26 & 832 \\ 
& 16 & 89.7\% & 243 & 7776 & 92.3\% & 90.7\% & 93 & 190 & 93.1\% & 92.4\% & 26 & 832 \\ 
& 24 & 92.6\% & 243 & 7776 & 94.9\% & 92.9\% & 227 & 544 & 95.3\% & 96.4\% & 32 & 1024 \\ 
& 32 & 91.9\% & 243 & 7776 & 93.3\% & 92.2\% & 243 & 1024 & 94.2\% & 95.9\% & 26 & 832
    \end{tabular}
    \caption{\label{tab:tQSTres-5} Same as before but for 5-qubit depth-3 random circuits.}
    \end{center}
\end{table*}

\begin{table*}[!b]
    \begin{center}
    \begin{tabular}{
        >{\centering\arraybackslash}m{10mm}| 
        >{\centering\arraybackslash}m{12mm}|| 
        >{\raggedleft\arraybackslash}m{15mm}| 
        >{\raggedleft\arraybackslash}m{10mm}| 
        >{\raggedleft\arraybackslash}m{9mm}|| 
        >{\raggedleft\arraybackslash}m{15mm}| 
        >{\raggedleft\arraybackslash}m{15mm}| 
        >{\raggedleft\arraybackslash}m{10mm}| 
        >{\raggedleft\arraybackslash}m{10mm}|| 
        >{\raggedleft\arraybackslash}m{15mm}| 
        >{\raggedleft\arraybackslash}m{15mm}| 
        >{\raggedleft\arraybackslash}m{10mm}| 
        >{\raggedleft\arraybackslash}m{10mm}  
    }
        \multicolumn{2}{c||}{} & \multicolumn{3}{c||}{\bf{fQST}} & \multicolumn{4}{c||}{\bf{tQST}} & \multicolumn{4}{c}{\bf{ECT-QST}} \\
        \cline{3-13}
        \hline
        $N$ & $\{\rho_{ii}\}\neq0$ & \centering$F^{\rho_{\mathrm{fQST}}}_{\rho_{\mathrm{target}}}$ & \centering$|S|$ & \centering$M$ & \centering$F^{\rho_{\mathrm{tQST}}}_{\rho_{\mathrm{fQST}}}$ & \centering$F^{\rho_{\mathrm{tQST}}}_{\rho_{\mathrm{target}}}$ & \centering$|S'_t|$ & \centering$M$ & \centering$F^{\rho_{\mathrm{ECT}}}_{\rho_{\mathrm{fQST}}}$ & \centering$F^{\rho_{\mathrm{ECT}}}_{\rho_{\mathrm{target}}}$ & \centering$|S'_t|$ & $M$\hspace{4mm}\mbox{} \\
        \hline\hline
        \multirow{8}{*}{6} 
& 8 & 92.9\% & 729 & 46656 & 92.7\% & 92.6\% & 137 & 292 & 95.2\% & 96.6\% & 32 & 2048 \\ 
& 9 & 82.3\% & 729 & 46656 & 80.6\% & 81.3\% & 91 & 168 & 86.9\% & 89.8\% & 25 & 1600 \\ 
& 12 & 92.9\% & 729 & 46656 & 93.0\% & 93.3\% & 141 & 336 & 95.0\% & 94.3\% & 27 & 1728 \\ 

& 16 & 90.0\% & 729 & 46656 & 90.9\% & 90.0\% & 259 & 586 & 93.9\% & 94.9\% & 36 & 2304 \\ 
& 24 & 87.9\% & 729 & 46656 & 90.3\% & 87.7\% & 535 & 1654 & 92.3\% & 92.8\% & 37 & 2368 \\ 
& 32 & 84.3\% & 729 & 46656 & 86.6\% & 84.3\% & 569 & 1830 & 88.2\% & 93.9\% & 41 & 2624 \\ 
& 48 & 87.0\% & 729 & 46656 & 88.3\% & 86.2\% & 565 & 1860 & 91.4\% & 93.7\% & 41 & 2624 \\ 
& 64 & 91.0\% & 729 & 46656 & 93.1\% & 90.9\% & 529 & 1306 & 94.4\% & 95.7\% & 42 & 2688 \\ 
    \end{tabular}
    \caption{\label{tab:tQSTres-6} Same as before but for 6-qubit depth-3 random circuits.}
    \end{center}
\end{table*}

\begin{table*}[!b]
    \begin{center}
    \begin{tabular}{
        >{\centering\arraybackslash}m{10mm}| 
        >{\centering\arraybackslash}m{12mm}|| 
        >{\raggedleft\arraybackslash}m{15mm}| 
        >{\raggedleft\arraybackslash}m{10mm}| 
        >{\raggedleft\arraybackslash}m{9mm}|| 
        >{\raggedleft\arraybackslash}m{15mm}| 
        >{\raggedleft\arraybackslash}m{15mm}| 
        >{\raggedleft\arraybackslash}m{10mm}| 
        >{\raggedleft\arraybackslash}m{10mm}|| 
        >{\raggedleft\arraybackslash}m{15mm}| 
        >{\raggedleft\arraybackslash}m{15mm}| 
        >{\raggedleft\arraybackslash}m{10mm}| 
        >{\raggedleft\arraybackslash}m{10mm}  
    }
        \multicolumn{2}{c||}{} & \multicolumn{3}{c||}{\bf{fQST}} & \multicolumn{4}{c||}{\bf{tQST}} & \multicolumn{4}{c}{\bf{ECT-QST}} \\
        \cline{3-13}
        \hline
        $N$ & $\{\rho_{ii}\}\neq0$ & \centering$F^{\rho_{\mathrm{fQST}}}_{\rho_{\mathrm{target}}}$ & \centering$|S|$ & \centering$M$ & \centering$F^{\rho_{\mathrm{tQST}}}_{\rho_{\mathrm{fQST}}}$ & \centering$F^{\rho_{\mathrm{tQST}}}_{\rho_{\mathrm{target}}}$ & \centering$|S'_t|$ & \centering$M$ & \centering$F^{\rho_{\mathrm{ECT}}}_{\rho_{\mathrm{fQST}}}$ & \centering$F^{\rho_{\mathrm{ECT}}}_{\rho_{\mathrm{target}}}$ & \centering$|S'_t|$ & $M$\hspace{4mm}\mbox{} \\        
        \hline\hline
        \multirow{14}{*}{7} 
& 16 & 72.8\% & 2187 & 279936 & 72.7\% & 68.4\% & 13 & 144 & 78.2\% & 77.6\% & 7 & 896 \\ 
& 18 & 87.0\% & 2187 & 279936 & 84.9\% & 86.4\% & 91 & 234 & 90.7\% & 91.0\% & 41 & 5248 \\ 
& 20 & 81.8\% & 2187 & 279936 & 77.6\% & 77.9\% & 65 & 202 & 84.5\% & 85.2\% & 28 & 3584 \\ 
& 24 & 84.1\% & 2187 & 279936 & 82.2\% & 83.2\% & 227 & 504 & 88.4\% & 90.3\% & 44 & 5632 \\ 
& 27 & 81.6\% & 2187 & 279936 & 75.9\% & 78.0\% & 109 & 264 & 85.9\% & 86.7\% & 33 & 4224 \\ 
& 28 & 86.5\% & 2187 & 279936 & 85.4\% & 84.9\% & 99 & 256 & 89.2\% & 91.3\% & 31 & 3968 \\ 
& 32 & 82.7\% & 2187 & 279936 & 82.0\% & 83.3\% & 805 & 2414 & 88.2\% & 93.5\% & 52 & 6656 \\ 
& 36 & 86.3\% & 2187 & 279936 & 84.8\% & 86.2\% & 511 & 1630 & 88.8\% & 92.4\% & 44 & 5632 \\  
& 42 & 81.7\% & 2187 & 279936 & 77.9\% & 78.4\% & 165 & 350 & 83.7\% & 86.6\% & 42 & 5376 \\ 
& 48 & 85.1\% & 2187 & 279936 & 84.4\% & 82.8\% & 325 & 646 & 89.8\% & 92.5\% & 50 & 6400 \\ 
& 64 & 88.8\% & 2187 & 279936 & 89.3\% & 88.6\% & 289 & 702 & 91.4\% & 91.5\% & 44 & 5632 \\ 
& 72 & 90.9\% & 2187 & 279936 & 90.9\% & 90.1\% & 451 & 1218 & 92.1\% & 92.6\% & 59 & 7552 \\  
& 96 & 86.2\% & 2187 & 279936 & 83.7\% & 83.7\% & 289 & 582 & 88.6\% & 89.5\% & 56 & 7168
    \end{tabular}
    \caption{\label{tab:tQSTres-7} Same as before but for 7-qubit depth-3 random circuits.}
    \end{center}
\end{table*}

\end{widetext}


%

\end{document}